\documentclass[aps,pra,reprint,showpacs,groupedaddress,floatfix]{revtex4-2}

\usepackage{multirow} 
\usepackage{url} 
\usepackage{graphicx} 
\usepackage{amsmath,amssymb}

\usepackage[caption=false]{subfig}    
\usepackage{stackengine}
 
\newcommand{\psit}{\widetilde \psi} 

\usepackage{soul}

\usepackage{color}
\definecolor{dred}{rgb}{0.80,0.00,0.00} 
\definecolor{dgreen}{rgb}{0.00,0.60,0.00} 
\definecolor{dblue}{rgb}{0.00,0.00,0.80} 
\definecolor{dmagenta}{rgb}{0.80,0.00,0.80}

\begin{document} 
 
\title{The Lamb shift in muonic hydrogen and the electric rms radius of the proton.} 
\thanks{The idea elaborated in this paper was outlined earlier in a less comprehensive paper \cite{Walcher:2012qp}.}

\author{Thomas Walcher}
\affiliation{Institut f\"ur Kernphysik, Johannes  Gutenberg-Universit\"at Mainz, 55128 Mainz, Germany.}

\date{April 12, 2023}  
 
\begin{abstract} 
The "proton radius puzzle" is the 7-standard-deviations difference of the charge radius of the proton as determined from the Lamb shift in electronic hydrogen and elastic electron scattering off the proton on the one side and the high precision determination from the Lamb shift in muonic hydrogen on the other side. 
So far the explanation of this difference has been mostly searched for in the limitations of the non-muonic experiments as the extrapolation to  $Q^2 \rightarrow 0$\,GeV$^2$ for electron scattering. Since the time scale of the vacuum polarization $e^+e^-$-pairs, causing the bulk contribution of the Lamb shift in muonic hydrogen, is very much shorter than that of the photon exchanges, causing the Coulomb interaction, it is argued that the muon on its orbit around the proton has to be considered as quasi particle dressed by  $e^+e^-$ pairs  and may not be treated as a bare particle in an external Uehling potential. The proper realization of this view makes the proton radius puzzle disappear. The value for the rms charge radius of the proton determined from the muonic Lamb shift, taking this distinction into account,  is $r_p = 0.87455(48)\,\textrm{fm}$ in agreement with the CODATA-2010 value.
\end{abstract}

 
\maketitle  

 
\section{Introduction}
\begin{figure}[ht]
\includegraphics[width=0.99\columnwidth]{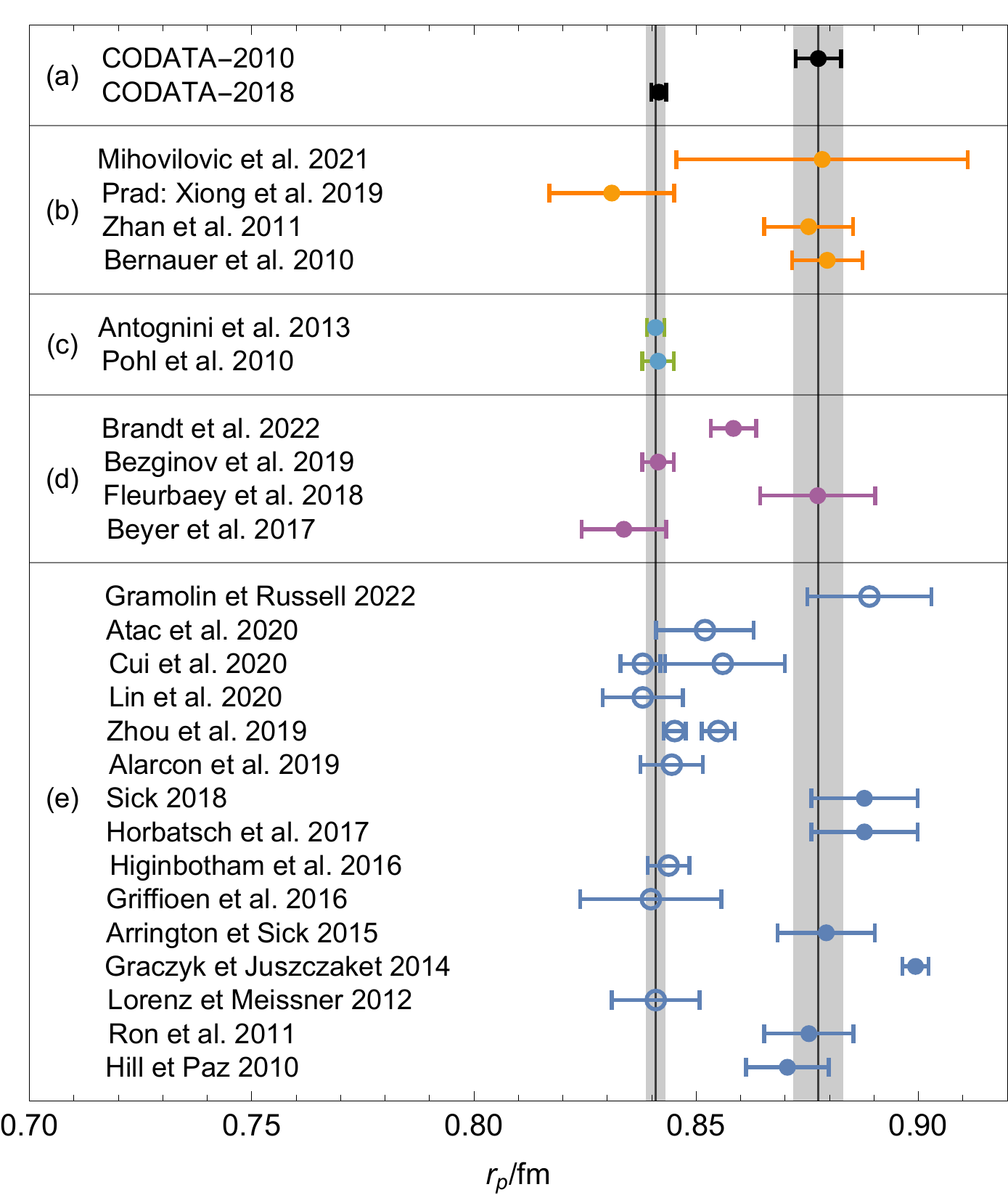}
\caption{Collection of $r_p$  values, see \cite{Gao:2021sml} for references. \\ (a) 7-std jump, (b) electron scattering, (c) muonic hydrogen (error bars $\times 5$), (d) electronic hydrogen, (e) reanalyses: mostly data from (b), open circles: extrapolation variations.}
\label{fig:1}
\end{figure}

About a decade ago a  7-standard-deviations difference of the root-mean-square (rms) charge radius of the proton $r_p = \langle r_p^2 \rangle^{1/2}$ derived from experiments with electrons and from the Lamb shift of muonic hydrogen was observed and had caused considerable attention in the physics community and beyond. Since the Lamb shift is a corner stone of the tests of Quantum Electrodynamics QED this difference requires indeed a convincing clarification. There was a hope that further experiments and their analysis would finally establish the correct value as summarized in the recent review paper of  Gao and Vanderhaeghen \cite{Gao:2021sml} and in Fig.\,\ref{fig:1}.  However, it shows that the values did not converge over time but still scatter around the  values of CODATA-2010 and CODATA-2018. 
The precise value of the rms radius derived from the Lamb shift in muonic hydrogen is $r_p = 0.84087(39)\,$fm \cite{Antognini:1900ns,Antognini:2012ofa} which has to be compared to the old CODATA-2010 value of 0.8775(51)\,fm \cite{Mohr:2012tt}, the 7-standard-deviations difference. The CODATA-2010 value is based on a measurement of the Lamb shift in electronic hydrogen \cite{Udem97} and on electron scattering experiments  \cite{Bernauer:2010wm,Zhan:2011ji,Ron:2011rd,Bernauer:2013tpr,Sick:2018fzn}.  Due to the excellent accuracy of the muonic experiment the reason for the deviation is mostly searched for in the electronic experiments having a lesser precision whether in electron scattering or in optical experiments with electronic hydrogen \cite{Gao:2021sml}. Therefore the new CODATA-2018 value of 0.8414(19)\,fm \cite{Tiesinga:2021myr} is based on a different weighting of the different experiments, however, there exist no good reasons to doubt and even to exclude the older experiments and the controversy  stays. 

With electron scattering experiments, the rms-radius is determined from the slope of the form factor as function of $Q^2$ at $Q^2 \rightarrow 0$\,GeV$^2$. In many attempts for explaining the deviation the validity of this extrapolation is questioned \cite{Meissner:2022rsm}, but this is already since some time incompatible with the detailed analysis of  Sick  \cite{Sick:2017aor,Sick:2018fzn}. 
Questioning the low $Q^2$ expansion of the form factor with its well defined and constrained parameters $\langle r^2 \rangle, \langle r^4 \rangle, \dots$, basically means questioning the validity of QED for this scattering \cite{Distler:2015rkm}. 

However, the calculation of the muonic Lamb shift is not as ironclad as it is mostly assumed. Weinberg (\cite{Weinberg:1995mt}, chpt. 11.2) argues, after his calculations of the radiative and vacuum polarization effects causing the Lamb shift in electronic atoms, that in muonic atoms one has to be careful by saying: "However, in this case the muonic atomic radius is not much larger than the electron Compton wave length, so the approximate result [in equation] (11.2.39) only gives the order of magnitude of the energy shift due to vacuum polarization." Though his rough equation has little to do with the sophisticated calculations \cite{Borie:1982ax,Pachucki:1996zza,Eides:2000xc,Borie:2004fv,Carroll:2011rv,Borie:2012zz}, used for the analysis of the muonic  measurement, the remark hints to an overlooked issue. 

Weinbergs conjecture can be extended by considering the order of magnitude parameters of the Bohr model of hydrogen. The Compton wave length of the $e^+e^-$ pair of the vacuum polarization is $\lambda_C = (\hbar c)/ (2 m_e c^2) \approx 193$\,fm, the muonic Bohr radius for $n=1$ $r_{\mu,\,B}  = (1/\alpha) (\hbar c)/(m_{\mu} c^2) = (2/\alpha) (m_e/m_{\mu}) \lambda_C \approx  256$\,fm. This means one has up to $(2 \pi r_{\mu,\,B}) / \lambda_C \approx 10$ separate and independent  spatial interaction regions for the muon mediated by the $e^+e^-$ pair fluctuations on its way around the proton.  From this follows that  according to quantum mechanical scattering theory the muon experiences consecutive independent interactions via $e^+e^-$ pairs. These interactions happen on a time scale of $\tau_C \approx 10^{-21}s$ which is close to zero compared to the time for one muon-orbit round of $t_{\textrm{round}} \approx 10^{-18}s$. One has to imagine the "asymptotic" in-out states in the scattering mediated by the $e^+e^-$ pairs, not as plane waves, but as the wave function of the bound muon in the external Coulomb field. The interactions are independent and produce a time-dependence of the interaction  on the mentioned scales.  

On the other hand the orbits of the muon are produced by the Coulomb interaction which one had actually to consider for consistency reasons also in the QED frame work. This interaction is equivalent to the exchange of many photons having infinite interaction ranges. Since no one knows how to put the sum of photon exchange diagrams in a wave equation used for the description of atoms, one just uses the external Coulomb potential, which can be justified by summing up all ladder diagrams of the multi-photon exchanges as has been shown by Weinberg (\cite{Weinberg:1995mt}, chpt. 13.6). This Coulomb potential is evidently time-independent in contrast to the potential one has to assign to the multi $e^+e^-$ pair exchanges for which the proof of Weinberg does not work. The Coulomb potential may be viewed as a time-independent external potential, whereas the $e^+e^-$-pair potential is time-dependent. The approximation used so far in the calculations of the muonic Lamb shift is the Uehling potential, due to the exchange of  $e^+e^-$ pairs, assumed to act on the same time scale as the photon interaction and consequently is taken to be time-independent \cite{Borie:1982ax,Pachucki:1996zza,Eides:2000xc,Borie:2004fv,Carroll:2011rv,Borie:2012zz}. This means that the $e^+e^-$ pair potential is taken also as an external potential just as the Coulomb potential in the wave equations, forgetting about its time dependence due to the series of separate scatterings. The Uehling effect is not a potential as a correction to the Coulomb potential as said in ref.\,\cite{Eides:2000xc} before their equation\,(200).

If one wants to take the correct approximations of the interaction in muonic atoms with the external Coulomb potential und the       time-dependent $e^+e^-$ pair potential one encounters the problem that one has to stay in the frame work of time-dependent wave equations since the time dependency cannot be separated in the usual way by assuming an exponential with a constant energy (\cite{Schiff:1968yy}, chpt. 26). This applies also for more sophisticated formulations of the scattering theory as e.g. with the Lippmann-Schwinger equation (\cite{Newton:1982qc}, chpt. 6 and 7). Fortunately, there exist solutions for an analog problem, namely the passing of electrons through a crystal: time-dependent corrections, described by Feynman diagrams, have to be applied to the solution with the time-independent potential created by the crystal-atoms.

Mattuck mentions in "A Guide to Feynman Diagrams in the Many-Body Problem" \cite{Mattuck:1976xt}  how the quantum mechanical "pinball" idea can be used to solve the problem not only for many body problems but also for atoms (\cite{Mattuck:1976xt}, chpt. 4.7, p.\,91). The analogy is in the many independent interactions, not in the many particles of the crystal or the one proton, producing the external potential confining the electrons or the one muon, respectively. In the following it will be shown how this works and that the modification of the Lamb shift calculation realizing time-dependent $e^+e^-$ pair exchange instead of using the time-independent external Uehling potential solves the proton-charge-radius problem.


\vspace{1cm}
\section{Basic theory \label{basic}} 

\begin{figure}[ht]
\centering
\includegraphics[width=0.4\columnwidth]{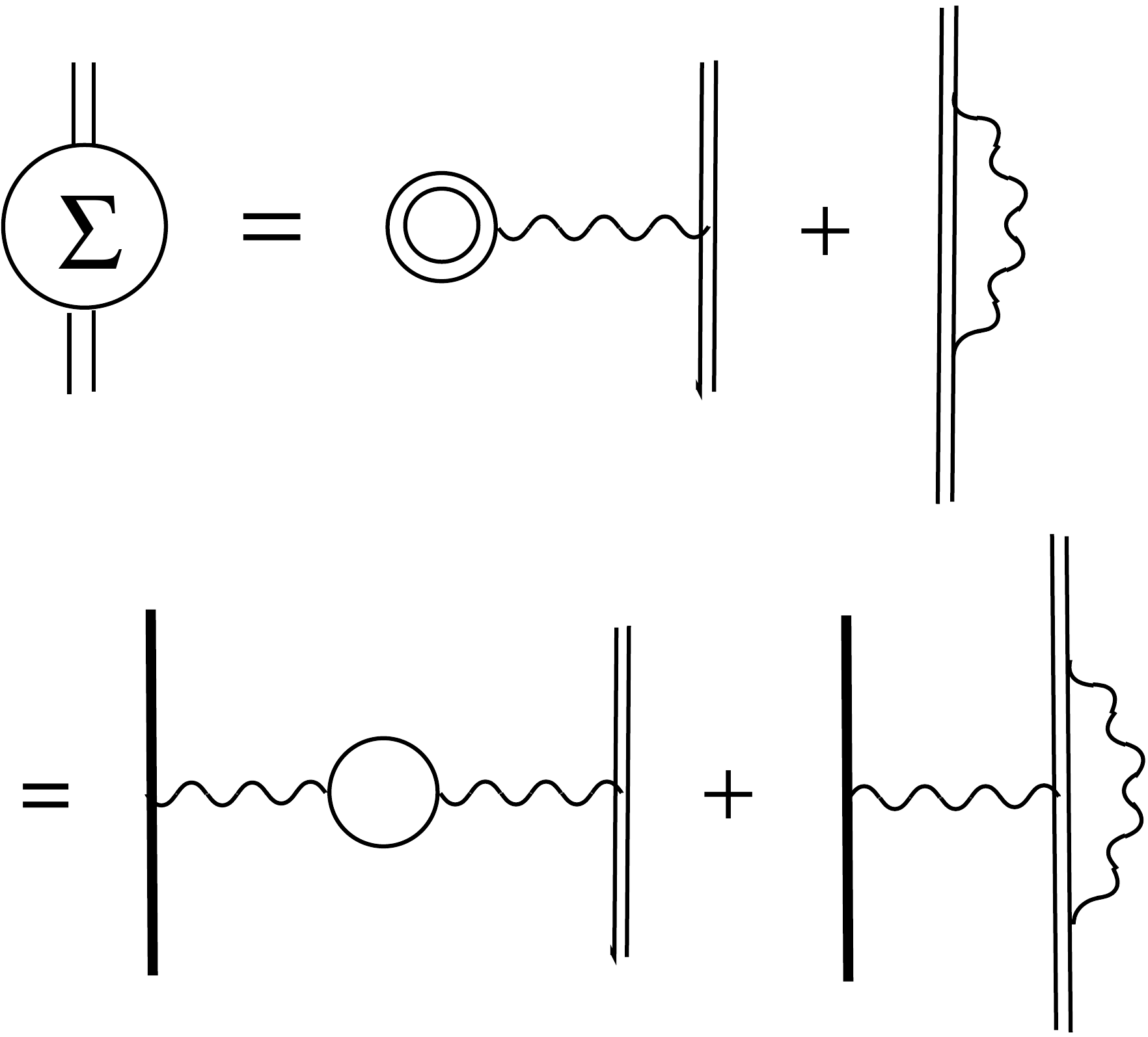}
\caption{The muon self energy $\Sigma$  consisting of radiative corrections and the vacuum polarization loop. Double lines are a short hand for the influence of the external Coulomb field on the propagators and the thick lines indicate the proton.}
\label{fig:2}
\end{figure}

\begin{figure}[ht]
\makebox[\columnwidth]{
\hspace{9cm}
\includegraphics[width=1.2\columnwidth]{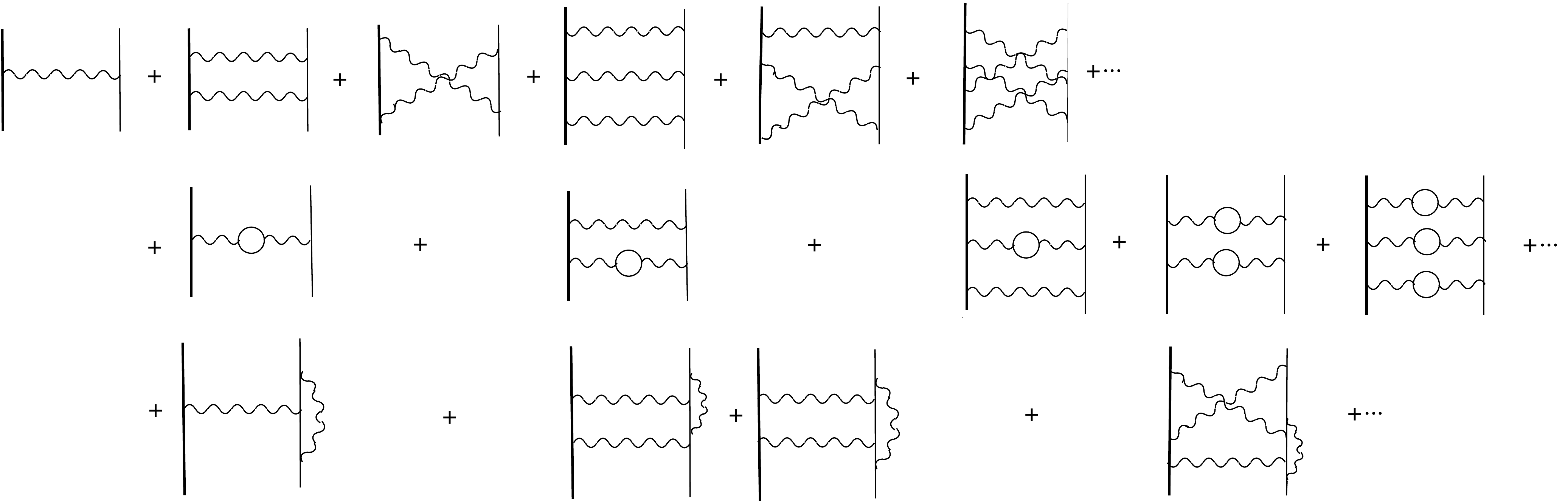}}
\parbox[c]{2 \columnwidth}{\caption{A selection of Feynman diagrams contributing to the interaction between muon and proton.}}
\label{fig:3}
\end{figure}

For a more detailed discussion of the arguments presented in the introduction we summarize some basic theory.  The Lamb shift is due to the "self-energy" diagrams, i.e. the radiative corrections and the vacuum polarization. For the electronic hydrogen the radiative corrections dominate and the vacuum polarization is small  whereas the reverse holds for muonic hydrogen (see e.g. Weinberg \cite{Weinberg:1995mt}, chpt. 14.3). The two diagrams of the self-energy interactions are depicted in Fig.\,\ref{fig:2}. 
The thick line indicates the proton and double lines are a short hand for the influence of the external Coulomb field on the propagators via photon exchange. 
The task for describing the muonic hydrogen correctly in the frame work of QED is now to sum up all possible Feynman diagrams allowed with the proper time ordering to all orders as shown in Fig.\,\ref{fig:3}. The sum of the diagrams had to be calculated via the sum of the G-Matrix iterations and its poles gave the energies of the bound states including all self-energy corrections. This approach is of course not feasible and one has to find an approximation reflecting the situation outlined. We first remark that the proton and the muon are not relativistic and only the ladder diagrams ordered by $L(\alpha^1)$, $L(\alpha^2)$, $L(\alpha^3)$, $\dots$ are important. As mentioned Weinberg proofs (\cite{Weinberg:1995mt}, chpt. 13.6) that the sum of the uncrossed photon ladder diagrams gives the familiar external Coulomb potential.   With "external" one signifies an interaction which is not represented by a Feynman diagram in analogy to e.g. the external electric field of the Stark effect. It does not depend on propagators or wave functions of the particles interacting. 
\onecolumngrid   
\twocolumngrid
The spin-orbit coupling is an automatic consequence of the Dirac theory  with the time-independent Coulomb potential and is therefore also an external interaction. The same is true for the hyperfine interaction.

Next one has to consider the radiative corrections which, however, are small in muonic hydrogen compared to the vacuum polarization \cite{Weinberg:1995mt}. The radiative corrections have the same range as the photon exchanges producing the Coulomb potential and therefore they are approximated as time-independent. This approximation might explain why some results with the electronic Lamb shift scatter. The radiative corrections are not further considered in this paper.

The diagrams containing $e^+e^-$ pairs, loops or "bubbles" \cite{Mattuck:1976xt} plus additional photons are at least by order $\alpha$ smaller than the ones containing the same number of  $e^+e^-$ bubbles and may be neglected.

\begin{figure}[ht]
\vspace{2mm}
\centering
\includegraphics[width=0.80\columnwidth]{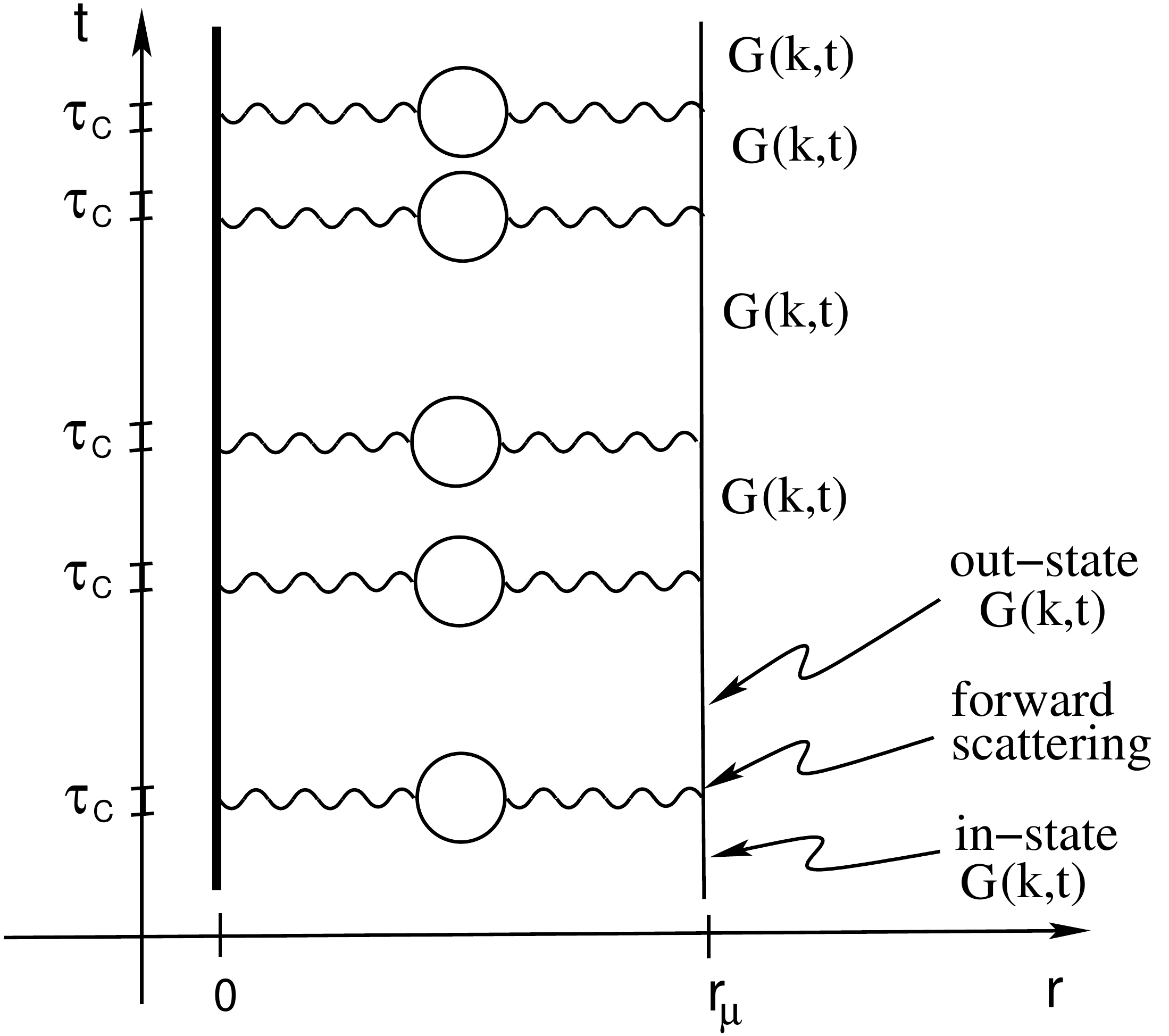}
\caption{Diagram of the pinball scattering in a bound state of muonic hydrogen. The forward scattering happens through in- and out-states with $G^-(k_{\textrm{in}},t) = G^+(k_{\textrm{out}},t) =  G(k,t)$. $\tau_C \approx 10^{-21}$\,s is the lifetime of the $e^+e^-$ pairs of the vacuum polarization and $\approx 0$\, compared to the characteristic time scales as required for interactions mediated by bubbles.}
\label{fig:4}
\end{figure}

The sum of the vacuum polarization ladder diagrams produce the major effect of the muonic Lamb shift and has to be treated differently than the sum of photon exchanges for the reasons sketched in the introduction. A more detailed picture is given in Fig.\,\ref{fig:4}. The  $e^+e^-$ loops have a life time of $\tau_C \approx 10^{-21}s$ and are independent one from the other. Consequently they will have to be summed up as products of probability amplitudes with $(k_{\textrm{in}},E) = (k_{\textrm{out}},E) $. These interactions represent "forward scattering" (see \cite{Mattuck:1976xt}, chpt. 4.5, after eq.\,(4.53) and p.\,82) just changing the unperturbed wave function to the perturbed one resulting in the momentum distribution of the bound state-wave function. 

Since the scattering state is a bound shell state the in-state wave function has to equal the out-state wave function on its closed orbit and, therefore, one may set $G^-(k_{\textrm{in}},E) = G^+(k_{\textrm{out}},E) = G(k,E)$ . We shall see in the next section how this peculiar state can be obtained. 

\begin{figure}[ht]
\vspace{2mm}
\centering
\includegraphics[width=0.70 \columnwidth]{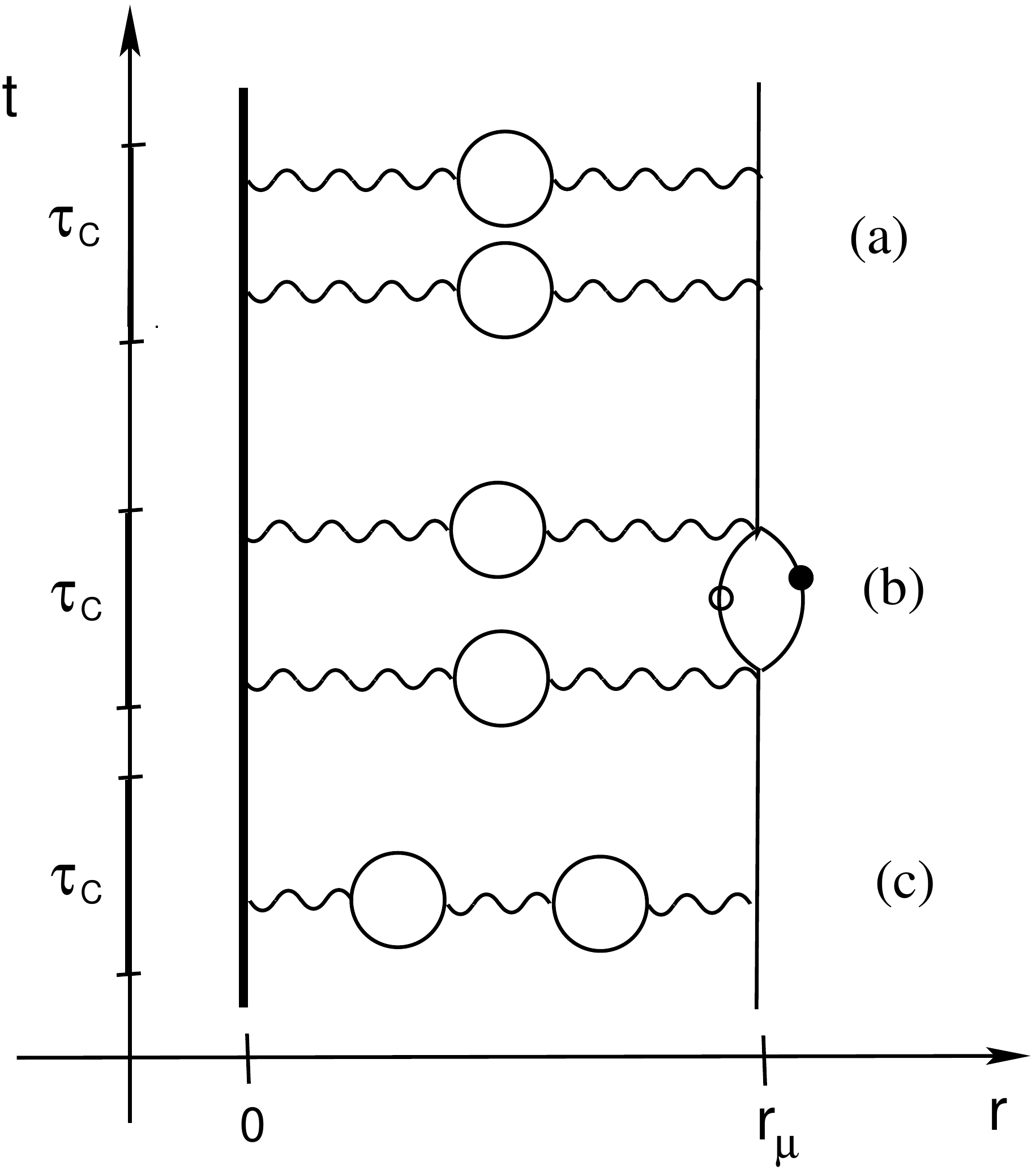}
\caption{Order $\alpha^4$ diagrams: (a) "double vacuum polarization correction" \cite{Pachucki:1996zza}, also "polarization insertions in two Coulomb lines" \cite{Carroll:2011rv}, (b) muon-hole excitations of the shell, (c) "two loop vacuum polarization" \cite{Pachucki:1996zza}.}
\label{fig:5}
\end{figure}

The diagrams of next to leading order $\alpha^4$ shown in Fig.\,\ref{fig:5} have to be distinguished from those of Fig.\,\ref{fig:4} since they are coherent within $\tau_C$.  Fig.\,\ref{fig:5}(a) and Fig.\,\ref{fig:5}(b) represent loops internally off-mass shell and their momenta have to be integrated and summed over.  But, as will be shown in section \ref{results}, the influence of the perturbed wave function is very small so that its change compared to the unperturbed wave function can be neglected.


\section{Theoretical details and approximations \label{approx}}

\begin{figure}[ht]
\centering
\subfloat[reiterated form]{\includegraphics[width=0.35\columnwidth]{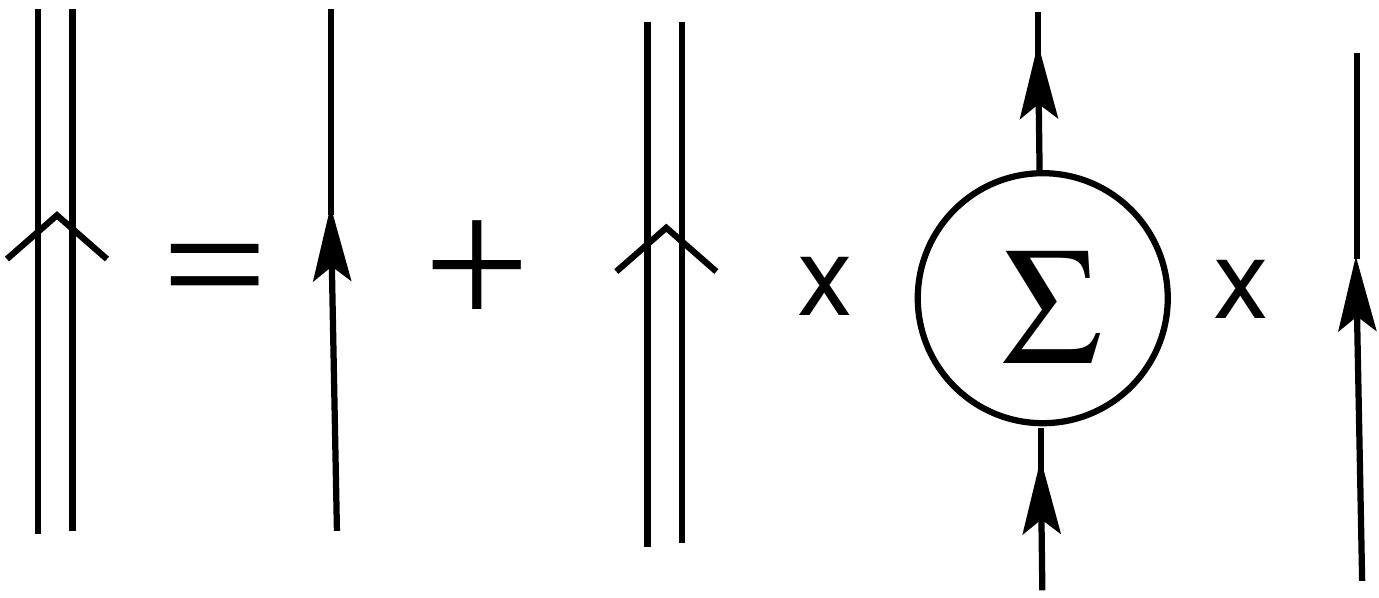}}
\caption{The generalized Dyson series in the presence of the external Coulomb potential in three different forms. (See Mattuck \cite{Mattuck:1976xt}, chpt.\,10.2  for the convention of the diagrams.)}
\label{fig:6}
\end{figure}

After outlining the basic idea one has to show how one can sum up all diagrams forming effectively a generalized Dyson series as depicted in Fig.\,\ref{fig:6}(a). We shall neglect the effect of the size of the proton which is a separate issue since the searched for difference is  in the Lamb shift for a point charge. This issue is discussed in the literature (see e.g \cite{Borie:1982ax,Pachucki:1996zza,Borie:2004fv,Carroll:2011rv,Borie:2012zz}). Due to the multiplicative property of the forward scattering amplitudes a geometric series emerges which can be summed up as symbolically shown in Fig.\,\ref{fig:6}(b).
\begin{figure}[ht]
\centering
\subfloat[expanded form]{\includegraphics[width=0.57\columnwidth]{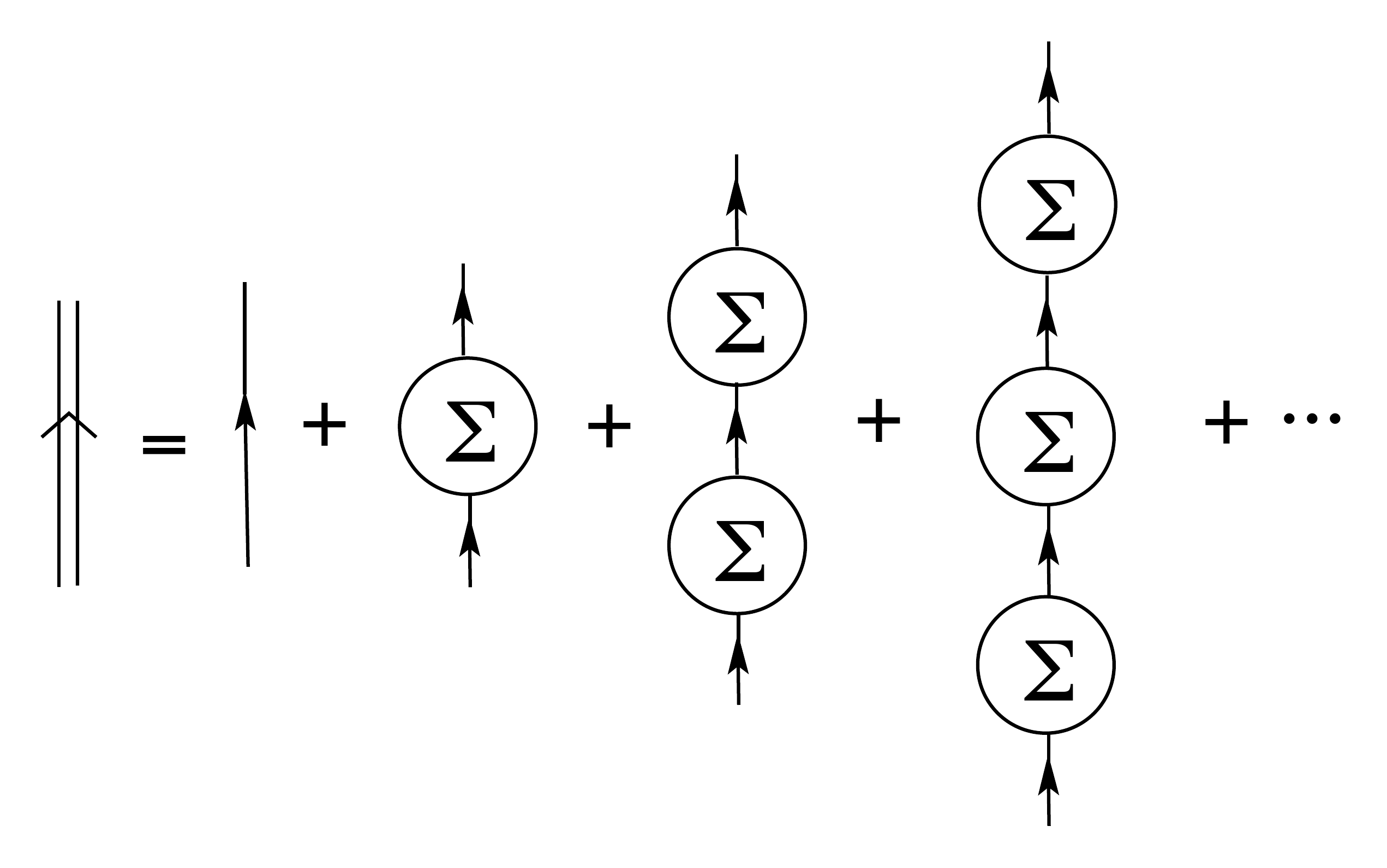}} 
\end{figure}
\begin{figure}[ht]
\centering
\subfloat[geometric sum]{\includegraphics[width=0.44\columnwidth]{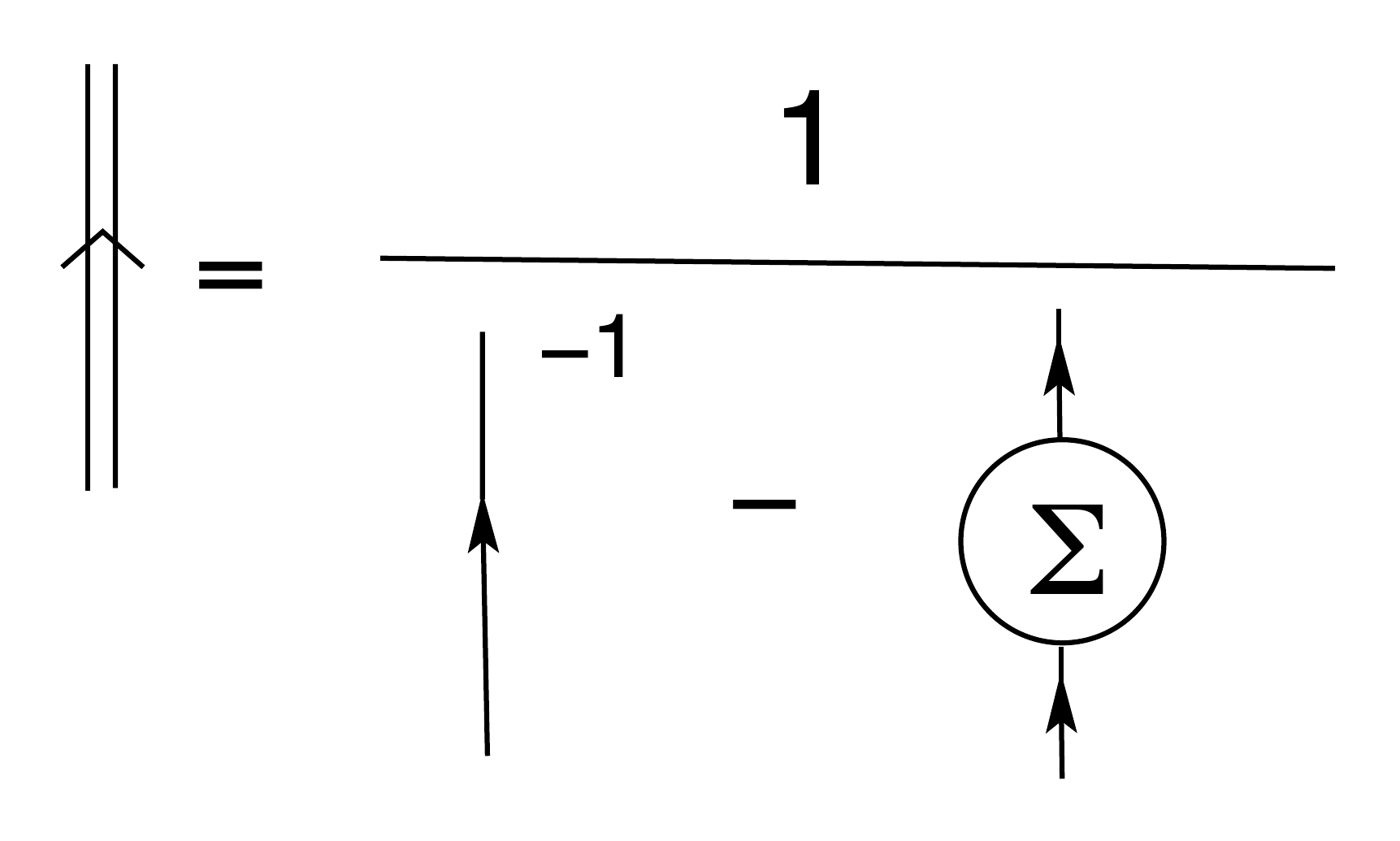}} 
\end{figure}
The diagram in Fig.\,\ref{fig:6}(b) can be transformed into the analytical form of the G-matrix:
\begin{equation}
G(\vec{k},E) =  \frac{1}{E - E^{(0)}_n - \langle \psit_n | \Sigma(\vec{k},E) | \psit_n \rangle} 
\label{eq:1}
\end{equation}
with $\langle  \psit_n |  \psit_n \rangle = 1$. The pole at
\begin{equation}
\Delta E_{LS} = \widetilde{E}_n - E^{(0)}_n = \langle \psit_n | \Sigma(\vec{k},\widetilde{E}_n) | \psit_n \rangle
\label{eq:2}
\end{equation}
represents the Lamb shift in leading order given by $\Sigma$ with the up to here unknown exact wave functions $ \psit_n$. In order to arrive at this formula we have integrated over the coordinates of the proton wave function being in good approximation a $\delta$-function only (see \cite{Mattuck:1976xt}, eq.\,(4.73)). 

The diagram in Fig.\,\ref{fig:6}(c) reads in the form of the G-matrix:
\begin{equation}
G(\vec{k},E) = G_0(\vec{k},E) + G_0(\vec{k},E) \, \Sigma(\vec{k},E) \, G(\vec{k},E)
\label{eq:3}
\end{equation}
as a Lippmann-Schwinger equation for forward scattering amplitudes. This equation has to be solved iteratively starting with the unperturbed $G_0(\vec{k},E)$ for $G(\vec{k},E)$ on the right hand side, calculating $G_1(\vec{k},E)$, inserting this again on the right hand side calculating $G_2(\vec{k},E)$ and continuing until nothing changes any more. Since eq.\,(\ref{eq:3}) is an operator equation it is easier to take advantage of the observation that it is fully equivalent to the Hartree-Fock equation (\cite{Mattuck:1976xt}  chpt. 11.1, \cite{Hedin:195aa}). Since we have only one muon orbiting we do not have to anti-symmetrize and the Hartree equation is appropriate.

 From our consideration above we know the Lamb shift as the self-energy eq.\,(\ref{eq:2}). Adapting the considerations of Mattuck (\cite{Mattuck:1976xt}, chpt. 4.7, pp.\,90-91) on Hartree quasi particles we arrive at the time-independent Schr\"odinger  equation:
 \begin{equation}
(T + V_{\textrm{Coulomb}})\, \psit_n + \langle \psit_n |\Sigma | \psit_n \rangle \psit_n = \widetilde{E}_n\,  \psit_n 
\label{eq:4}
\end{equation}
The $\psit_n$ are the exact solutions of the Hartree equation with the condition $\langle  \psit_n |  \psit_n \rangle = 1$.  This equation is solved iteratively until a self consistent solution $\psit_n$ is achieved. This procedure is called "self-consistent renormalization in an external field" (\cite{Mattuck:1976xt} chpt. 11.1).
The self-consistent procedure implied by eq.\,(\ref{eq:4}) can be easily performed in the following way. We begin with the unperturbed wave function $\psi^{(0)}_n$ and the usual ansatz:
\begin{equation}
\psi_n = \psi^{(0)}_{n} + \delta \psi_n,  ~  \delta \psi_n \bot \psi^{(0)}_n ~ \textrm{and}  ~ \langle \psi^{(0)}_n | \psi^{(0)}_n \rangle =1.
\label{eq:5}
\end{equation}
this means that no orthonormal basis for expanding $\psi_n,$ as in the time-independent perturbation theory is chosen.  Since $\psi_n$  is not normalized we set  $\psit_n = \psi_n / \langle \psi_n | \, \psi_n \rangle ^{1/2}$ and eq.\,(\ref{eq:4}) reads
\begin{equation}
(T + V_{\textrm{Coulomb}})\, \psi_n + \frac{\langle \psi_n |\Sigma | \psi_n \rangle}{\langle \psi_n | \, \psi_n \rangle} \, \psi_n = \widetilde{E}_n\,  \psi_n .
\label{eq:6}
\end{equation}

Multiplying from left with $\langle \psi^{(0)}_n |$ one gets an integral equation which lends itself well for the iterative solution:  
\begin{equation}
 \frac{\langle \psi^{(0)}_n | T + V_{\textrm{Coulomb}} |\, \psi_n \rangle}{ \langle \psi^{(0)}_n | \psi_n \rangle} +  \frac{\langle \psi_n |  \Sigma | \psi_n \rangle}{\langle \psi_n | \psi_n \rangle}  = \widetilde{E}_n 
 \label{eq:7}
 \end{equation}
From this equation follows in accord with eq.\,(\ref{eq:2}) derived for normalized wave functions: 
\begin{equation}
\Delta E_{n \,LS} = \widetilde{E}_n - E^{(0)}_{n} =  \frac{\langle \psi_n | \Sigma | \, \psi_n \rangle}{\langle \psi_n | \, \psi_n \rangle}.
\label{eq:8}
\end{equation}

If one dismisses the request that the Lamb shift is the self-energy and identifies the operator $\Sigma$ with an external potential we get a wave equation with the external "Uehling potential" $V_{\textrm{Uehling}}$ \cite{Borie:1982ax,Pachucki:1996zza,Borie:2004fv,Carroll:2011rv,Borie:2012zz}.  This is just the step from the "self energy" point of view to the "external potential" point of view. The solution of this wave equation gives somewhat different $\psi_n$ and $E_n$, denoted by $\psi^{\prime}_n$ and $E^{\prime}_n$:
 \begin{equation}
(T + V_{\textrm{Coulomb}})\, \psi^{\prime}_n +  V_{\textrm{Uehling}}\, \psi^{\prime}_n = E^{\prime}_n\,  \psi^{\prime}_n.
\label{eq:9}
\end{equation}
If one solves this wave equation numerically with and without  $V_{\textrm{Uehling}}$ the eigenvalues are independent of the normalization of $\psi^{\prime}_n$, and $\Delta E^{\prime}_{n \,LS} $ can be calculated directly \cite{Carroll:2011rv}.

However, more insight is gained if one multiplies eq.\,(\ref{eq:9}) with $\langle \psi_n^{(0)} |$ from the left. One obtains for the Lamb shift in this approximation:
\begin{equation}
\Delta E^{\prime}_{n\;LS} = E^{\prime}_n - E^{(0)}_n = \frac{ \langle \psi_n^{(0)} | V_{\textrm{Uehling}}\, | \psi^{\prime}_n \rangle}{\langle \psi_n^{(0)} | \psi^{\prime}_n \rangle}.
\label{eq:10}
\end{equation}
Eq.\,(\ref{eq:10}) differs from eq.\,(\ref{eq:8}) by the appearance of the unperturbed wave function  $\langle \psi_n^{(0)} |$ instead of the perturbed  $\langle \psi^{\prime}_n |$ on the left side of the matrix elements. It is this difference which is responsible for the fact that the proton radius determined from muonic hydrogen deviates from that determined with electron scattering.

One may now ask what the difference of eq.\,(\ref{eq:10}) to the classic perturbation theory used in the key calculations of \cite{Borie:1982ax,Pachucki:1996zza,Borie:2004fv,Borie:2012zz} is. One has for increasing order of the expansion parameter $\lambda$ \cite{Sakurai:2011zz}:
\begin{align}
&{\cal{O}}(\lambda^1):     \Delta_n^{(1)}= \langle \psi_n^{(0)} | V | \psi_n^{(0)} \rangle   \nonumber \\
&{\cal{O}}(\lambda^2):      \Delta_n^{(2)}= \langle \psi_n^{(0)} | V | \psi_n^{(1)} \rangle   \nonumber \\
&{\cal{O}}(\lambda^3):      \Delta_n^{(3)}= \langle \psi_n^{(0)} | V | \psi_n^{(2)} \rangle  \nonumber \\
 &\cdots
 \label{eq:11}
\end{align}
where  $V = \Sigma$  is the perturbing potential, $| \psi_n^{(0)} \rangle$ the ket of the unperturbed state and the next kets of the indicated order. It is evident that $\sum_{i=0}^m | \psi_n^{(i)} \rangle \rightarrow | \psi^{\prime}_n \rangle$ for $m \rightarrow \infty$ and in this way one realizes that $\sum_{i=1}^m \Delta_n^{(i)} \rightarrow \Delta E^{\prime}_{n\;LS}$. So the higher order energy shift of time-independent perturbation theory converges to the value of the numerical solution of eq.\,(\ref{eq:9}). 

 In time-independent perturbation theory, which takes the Uehling potential as an external potential, one expands in the basis of unperturbed eigenstates and gets:
 \begin{multline}
 \Delta E_{n \; LS} = \langle \psi^{(0)}_n | \Sigma | \psi^{(0)}_n \rangle \; + \\
    \sum_{k \neq n} \int_{E_k \geqslant 0} \frac{ \langle \psi^{(0)}_n | \Sigma | \psi^{(0)}_k \rangle  \langle \psi^{(0)}_k | \Sigma | \psi^{(0)}_n \rangle}{E^{(0)}_n-E^{(0)}_k}.
 \label{eq:12}
 \end{multline}
This equation reminds that the perturbed state $| \psi^{(1)}_n \rangle$ is interpreted as the sum over intermediate excitations 
shown in  Fig.\,\ref{fig:5}(b) as part of the time-dependent perturbation theory. So these intermediate excitations also appear in the external potential approximation for $\Sigma$ and the numerical solution of eq.\,(\ref{eq:9}) contains them through $| \psi^{\prime}_n \rangle$. One can account for these higher order contributions either summing over the intermediate states of eq.\,(\ref{eq:12}) or by solving eq.\,(\ref{eq:9}) numerically.

The reference point of the calculations of  \cite{Borie:1982ax,Pachucki:1996zza,Borie:2004fv,Borie:2012zz} is the first order term in eq.\,(\ref{eq:12}), i.e. first order Lamb shift, to which all corrections are added. The most important correction due to the second term $\Delta_n^{(2)}$ is the "double vacuum polarization correction" \cite{Pachucki:1996zza}. We shall  use this reference point as well and add our correction determined from the solution of eq.\,(\ref{eq:8}) to it. It is important to realize that these solutions contain no higher order contributions as they are shown in Fig.\,\ref{fig:5}. The pinball mechanism does not allow these higher order coherent diagrams.  Therefore, no higher  order contributions equivalent to the expansion in eq.\,(\ref{eq:11}) are produced by solving eq.\,(\ref{eq:7}) numerically. This means that the $\Delta E_{n\;LS}$ in eq.\,(\ref{eq:8}) does not contain the "double vacuum polarization correction". 

The higher order contributions and further corrections have been discussed and calculated in detail in \cite{Borie:1982ax,Pachucki:1996zza,Pachucki:1999zza,Borie:2004fv,Borie:2012zz} and the question arrises whether one has to redo them in the self consistency framework used here. However, as will be seen in the next section  \ref{results},  the self consistent wave functions change only very little compared to the unperturbed ones, just enough to explain the small Lamb shift difference producing the radius difference. Therefore, one may use for higher order corrections the unperturbed wave functions and stay with the corrections available in the literature. In principle one should include the coherent higher order contributions as in Fig.\,\ref{fig:5} as part of the $\Sigma$ to the geometric sum of the Dyson series and then solve eq.\,(\ref{eq:7}) self consistently. This is actually done in many body problems \cite{Mattuck:1976xt}, but is not needed here. 

On the other hand Carroll et al. \cite{Carroll:2011rv} solved the adaption of eq.\,(\ref{eq:9}) as a time-independent Dirac equation  with the external Uehling potential numerically, effectively including the higher order contributions of the time-independent perturbation theory according to $\sum_{i=1}^m \Delta_n^{(i)} \rightarrow \Delta E^{\prime}_{n\;LS}$. In this way the "double vacuum polarization correction" is already included in their numerical solution of the wave equation. Therefore, they correctly did not add this correction ("polarization insertion in two Coulomb lines" \cite{Carroll:2011rv}) to the reference point. This point will be discussed further at the end of section \ref{results}.

The time-dependent and the time-independent  perspectives are very different and reflect different physics ideas. However, it should be  clear that no double counting is done.

As shown with the derivation of  eq.\,(\ref{eq:2}), one has to use eq.\,(\ref{eq:8}) for calculating  $ \Delta E_{n \,LS}$ with the self-consistent renormalization which can only be done by iteration. As the first step of the iteration we solve eq.\,(\ref{eq:9}) which is according to eq.\,(\ref{eq:10}) already a good estimate. The resulting $\psi^{\prime}_n$ we insert as the first approximation $\psi_n^{(1)}$ for $\psi_n$  in eq.\,(\ref{eq:8}) yielding $\Delta E_{n \,LS}^{(1)}$. It turns out that this is already good enough. In order to show this we calculated the second step of the iteration by again using a perturbative ansatz:
\begin{equation}
\psi^{(2)}_n = \psi^{(1)}_{n} + \kappa \, \delta \psi^{(1)}_n,  ~  \delta \psi^{(1)}_n \bot \psi^{(1)}_n   
\label{eq:13}
\end{equation}
where the parameter $\kappa$ is varied until eq.\,(\ref{eq:7}) is satisfied with $\widetilde{E}_n  = E^{(1)}_n$. We do not assume that $\psi^{(1)}_{n}$ is normalized. The perpendicular change  $\delta \psi^{(1)}_n$ can be gained from:
\begin{equation}
\delta \psi^{(1)}_n = \psi_n^{(0)} - \frac{\langle \psi_n^{(0)} | \psi_n^{(1)} \rangle}{\langle \psi_n^{(1)} | \psi_n^{(1)} \rangle} \psi_n^{(1)}
\label{eq:14}
\end{equation}
where the choice of $\psi_n^{(0)}$ is somewhat arbitrary. One could have taken any Hilbert vector, but for a fast convergence the choice of $\psi_n^{(0)}$ is advantageous. The $\psi^{(2)}_n$ derived in this way is inserted in eq.\,(\ref{eq:8}) giving  $\Delta E_{n \,LS}^{(2)}$ which one compares to $\Delta E_{n \,LS}^{(1)}$.  It proves that the second iteration is by far good enough considering the errors of the radius determination from the muonic hydrogen.    


\section{Results \label{results}}

In distinction to Carroll et al. in ref.\,\cite{Carroll:2011rv}  who dealt with the relativistic Dirac equation we restrict ourselves to the non relativistic Schr\"odinger equation with the external Coulomb potential of a point charge. The Schr\"odinger equation  suffices for showing the correction of the Lamb shift without considering fine structure splitting, finite size effect, etc..  Relativistic effects on the Lamb shift are small \cite{Borie:2004fv} and do not change anything for this discussion. As the first step we calculate the exact solutions of the Schr\"odinger equation with the external point Coulomb potential and the external "Uehling potential"  according to  eq.\,(\ref{eq:9}) and then use eq.\,(\ref{eq:8}) for calculating the Lamb shift.

A compact representation of the "Uehling potential" $V_{\textrm{Uehling}}$, well suited for our calculation, is the representation provided by Pachucki \cite{Pachucki:1996zza}. 
(We have to be careful distinguishing between the "Uehling interaction" as the effect of the $e^+e^-$ bubbles on the muon as a "quasi particle", i.e. its "self energy",  and the "Uehling interaction" taken as an external potential. Since the analytical form is the same we shall use the notion "Uehling potential" for both situations. The respective  meaning is clear from the context.) 
For the numerical calculations of the Schr\"odinger equation we have used Mathematica. As Carroll et al. \cite{Carroll:2011rv} we have made extensive tests to guaranty the quality of the solutions. All calculations have been made with an internal precision of 64 digits and an accuracy goal of 20 digits. The optimal method is the "Explicit Runge-Kutta" integration for the S-State, and for the P-state the change between various methods provided "Automatic". The numerical eigenvalues of the unperturbed $2S$ and $2P$ states, i.e. without the "Uehling potential", are compared to the non relativistic exact solution (Bohr energies) and found to be good to a few neVs for different boundary conditions at small ($\approx 0.1$\,fm) and large ($\approx 10000$\,fm) radii. Of course, since we do not take the difference of the large energy eigenvalues as Carroll et al. \cite{Carroll:2011rv} but calculate the Lamb shift with eq.\,(\ref{eq:8}), we do not really need this extreme accuracy. 

When using our numerical wave functions $\psi^{(0)}$ from the numerical integration of eq.\,(\ref{eq:9}) without the perturbation $V_{Uehling}$, i.e. the wave equation for unperturbed states, and insert them in eq.\,(\ref{eq:8}) we get $\Delta E_{Lamb}  = 205.00463502$\,meV in complete agreement with the calculation using the exact analytical wave functions \cite{Pachucki:1996zza}. Additionally, as a further check, we have used the values $E_n$ and $E^{(0)}_n$, both obtained from the numerical integration of eq.\,(\ref{eq:9}) with and without $V_{Uehling}$, we get $\Delta E_{Lamb}  = E_n - E^{(0)}_n = 205.00463506$\,meV. This value for the proton with point charge is the mentioned reference point and the analyses of the muonic Lamb shift are resting on it so far. 

\begin{figure}[ht]
\centering
\includegraphics[width=0.95\columnwidth]{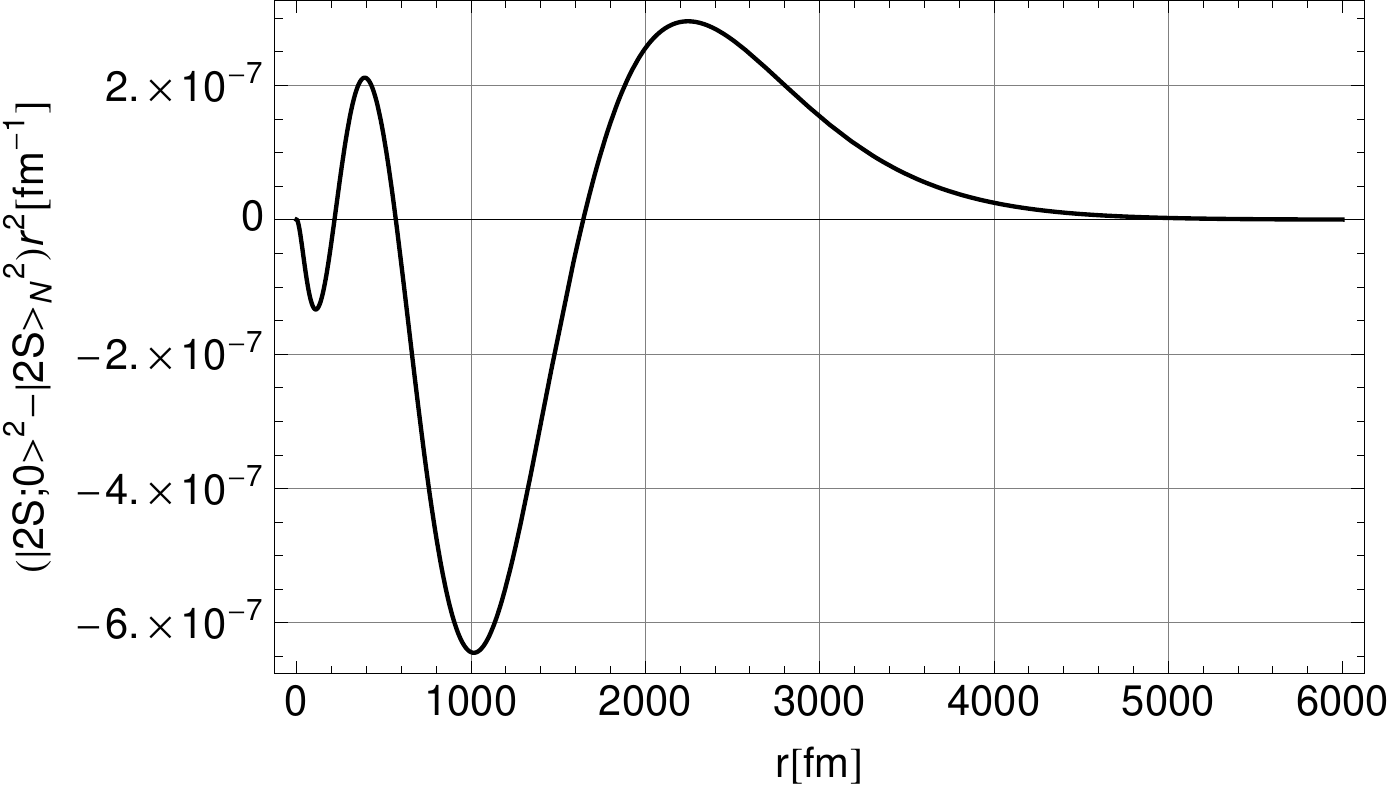}
\caption{The difference $(| 2\,S; 0 \rangle^2 -  | 2\,S \rangle_N^2) r^2$ showing the polarization charge density due to the "Uehling effect" for the $2\,S$ state divided by the negative elementary charge of the muon .}
\label{fig:7}
\end{figure}

\begin{figure}[ht]
\centering
\includegraphics[width=0.95\columnwidth]{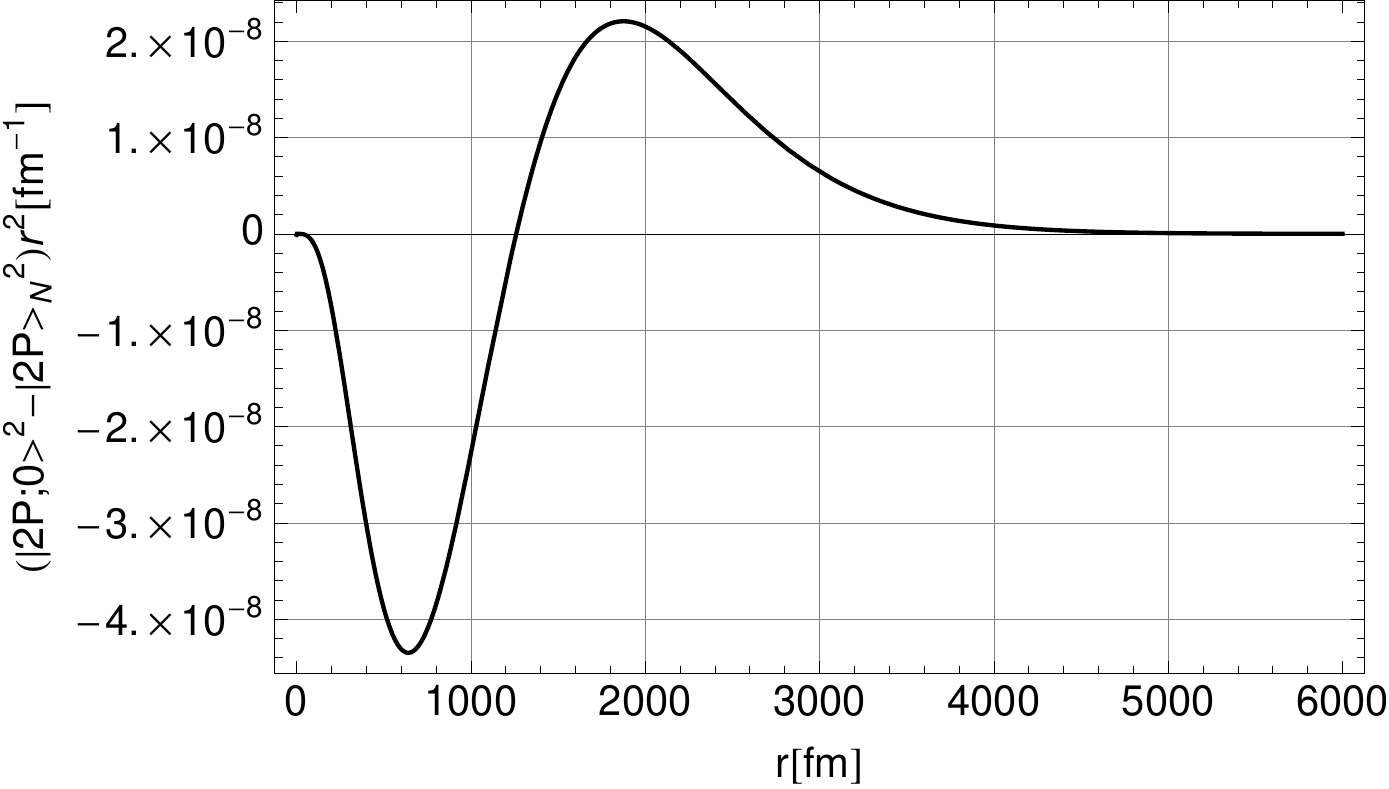}
\caption{The difference $ (| 2\,P; 0 \rangle^2 -  | 2\,P \rangle_N^2) r^2$ showing the polarization charge density due to the "Uehling effect" for the $2\,P$ state divided by the negative elementary charge of the muon }.
\label{fig:8}
\end{figure}

Figure \ref{fig:7} and Fig. \ref{fig:8} show the difference of the density of the unperturbed state $| n\,l ; 0 \rangle^2 r^2 = (\psi^{(0)}_{n\,l})^2 r^2$ minus the density of the normalized perturbed state $| n\,l  \rangle_N^2 r^2 = (\psi_{n\,l})^2 r^2 / \langle \psi_{n\,l} | \psi_{n\,l} \rangle$ for $n\,=\,2$ and $ l\,=\,0$, 1, respectively. 
One nicely sees the polarization charge induced in the vacuum by the muon. The wiggles at small radii in Fig. \ref{fig:7} are no numerical artifacts but due to the double bump structure of the 2S state.  In order to get an idea of the scales we note that the Bohr radius for the muon is $r_{\mu,\,B} = 268$\,fm, the scale of the Compton wave length of the electron-positron pair of the vacuum polarization $\lambda_C = 193$\,fm, the rms radius of the 2S state $\langle2\,S | r^2 |2\,S \rangle^{1/2} = 1854$\,fm and of the 2P state  $\langle2\,P | r^2 | 2\,P\rangle^{1/2} = 1560$\,fm. As expected some positive charge is pushed to larger radii compensated by negative charge at small radii indicating the induction of the polarization cloud in the vacuum. The argument of the importance considering the interaction scale  in the introduction by comparing just to the Bohr radius is amplified by realizing the about 5 times larger radii of the $n=2$ orbits.

Calculating the difference of the Lamb shifts with the perturbed $\psi^{(2)}_{2S}$ and $\psi^{(2)}_{2P}$ states according to  eq.\,(\ref{eq:8}) one gets the salient result of this paper:
\begin{equation}
\Delta E^{(2)\,\textrm{point charge}}_{2P \rightarrow 2S}  = E^{(2)}_{2P} - E^{(2)}_{2S} =  205.30658(100) \textrm{meV}
\label{eq:15}
\end{equation}
where the error indicates the variation of the value with different integration boundaries. The first iteration as described in the previous section yields:
\begin{equation}
\Delta E^{(1) \,\textrm{point charge}}_{2P \rightarrow 2S}  = E^{(1)}_{2P} - E^{(1)}_{2S} =  205.30664(100)\, \textrm{meV}
\label{eq:16}
\end{equation}

The result differs from the reference point for the unperturbed wave functions $\Delta E^{(0)\,\textrm{point charge}}_{\text{Lamb}} = 205.005(1)$\,meV by
\begin{equation}
\delta (\Delta E_{2P \rightarrow 2S})  = 0.302(1)\,\textrm{meV}.
\label{eq:17}
\end{equation}

The radius of the muonic hydrogen is derived from a combined measurement of the singlet transition $2\,S_{1/2}^{F=0} \rightarrow 2\,P_{3/2}^{F=1}$ and the triplet transition $2\,S_{1/2}^{F=1} \rightarrow 2\,P_{3/2}^{F=2}$. From these two measurements one can deduce both, the Lamb shift and the 2S-HF splitting independently \cite{Antognini:1900ns,Antognini:2012ofa}. From this determination of the Lamb shift one obtains the rms radius by equating it to the theoretical shift:
 \begin{multline}
 \Delta E^{\textrm{theory}}_{2P \rightarrow 2S} = (206.0336(15) - 5.2275(10)\,r_p^2 / \textrm{fm}^2 + \\ 
 0.0332(20))\,\textrm{meV}
 \label{eq:18}
\end{multline}   
The first term contains the Lamb shift with a multitude of corrections whereas the following terms represent the finite size effect due to the position probability of the muon at the proton \cite{Friar:1978wv,Pachucki:1996zza,Antognini:1900ns}. Some details of the finite size dependence of the Lamb shift  \cite{Carroll:2011rv} are neglected in \cite{Antognini:1900ns}. For comparability we follow \cite{Antognini:1900ns}. Using the two different radii given in the introduction and their errors as well as those of the constants in the theoretical formula given above one gets for the deviation of the Lamb shift:
\begin{equation}
\delta (\Delta E_{2P \rightarrow 2S}) = 0.329(50)\,{\textrm{meV}}
\label{eq:19}
\end{equation}
where the error is dominated by the electronic experiments. This result is in very good agreement with the result of eq.\,(\ref{eq:17}) and explains the "radius puzzle".

For the difference of the eigenvalues according to eq.\,(\ref{eq:10}) calculated with the numerical solution of the wave equation eq.\,(\ref{eq:9}) we get $\Delta E^{\,\textrm{point proton}}_{2P \rightarrow 2S} = 205.156(1)\,\textrm{meV}$. If we take the difference of the eigenvalues directly we get $\Delta E^{\,\textrm{point proton}}_{2P \rightarrow 2S} = 205.159(3)\,\textrm{meV}$. 
Since we do not use the difference of the eigenvalues we have not insisted to improve this error. Carroll et al. \cite{Carroll:2011rv} get for this difference  $205.1706(5)\,\textrm{meV}$ from the numerical solution of the Dirac equation. According to \cite{Pachucki:1999zza,Veitia:2004zz} we have to add a relativistic correction of 0.0169\,meV to our non-relativistic result  yielding $\Delta E^{\,\textrm{point proton}}_{2P \rightarrow 2S} = 205.172(1)\,\textrm{meV}$ in good agreement with the relativistic result of \cite{Carroll:2011rv}.


\section{Conclusions}
If the Lamb shift of a point charge is taken as the dynamical QED effect due to the interaction of the muon with itself via the vacuum polarization, its self-energy, and not as a shift caused by the "Uehling potential" taken as an time-independent external  potential, one gets agreement for the radius determined from the Lamb shift in muonic hydrogen and the combined electronic experiments. 

The finite size effect as given in eq.\,(\ref{eq:18}) is practically not modified by our considerations. The other corrections used for eq.\,(\ref{eq:18}), in particular the "double vacuum polarization" of 0.151\,meV, are to sufficient approximation not changed by the modified muon wave functions proposed here and we assume that they are the same as used in the analysis of Antognini et al. \cite{Antognini:1900ns}.  If we correct the Lamb shift of the point charge in eq.\,(\ref{eq:18}) with the difference of  eq.\,(\ref{eq:17}) we arrive at a new value for the rms radius of the proton derived from the muon experiment of Antognini et al. \cite{Antognini:1900ns}:
\begin{equation}
r_p = 0.87455(48)\,\textrm{fm}
\label{eq:20}
\end{equation}
which includes the error of the correction in eq.\,(\ref{eq:17}). This value is in good agreement with the best electron scattering rms radius  0.879(8)\,fm \cite{Bernauer:2010wm,Bernauer:2013tpr,Sick:2018fzn} and that of CODATA-2010 0.8775(51)\,fm \cite{Mohr:2012tt}.
For a final result for the radius derived from muonic hydrogen one had to redo the relativistic calculations including the finite size effects as  e.g. Carroll et al. \cite{Carroll:2011rv} realizing the considerations of this paper.  

\begin{acknowledgments}
The author is indebted to Joerg Friedrich and Volker Metag for a critical reading of the manuscript and for encouragement, and to Michael Distler for help with Fig.\,1 and beyond.
\end{acknowledgments}

\bibliographystyle{apsrev4-2} 
\bibliography{references_2023}
 
\end{document}